\documentclass[aps,prd,twocolumn,groupedaddress,showpacs,floatfix]{revtex4}
\usepackage{amssymb}
\usepackage{amsmath}
\usepackage{graphicx}
\begin{document}
\title{Asymptotic cosmological solutions for string/brane gases
       with solitonic fluxes}
\author{Antonio Campos}
\affiliation{Institut f\"ur Theoretische Physik,
             Universit\"at Heidelberg,
             Philosophenweg 16,
             69120 Heidelberg,
             Germany}
\date{\today}

\begin{abstract}
We present new cosmological solutions for brane gases
with solitonic fluxes that can dynamically explain the 
existence of three large spatial dimensions.
This reasserts the importance of fluxes for understanding 
the full space of solutions in a potential implementation 
of the Brandenberger-Vafa mechanism with M2-branes.
Additionally, we study a particular example in which
the cosmological dynamics supported by a string gas
with a NS flux in the ten-dimensional dilaton gravity 
framework is asymptotically equivalent to that of a M2-brane 
gas with a certain wrapping configuration in eleven-dimensional
supergravity.
We speculate that this connection between the ten- and 
eleven-dimensional implementations of the Brandenberger-Vafa
mechanism could be a general feature.
\end{abstract}
\maketitle

\section{Introduction}
In the late '80s Brandenberger and Vafa suggested a 
potential mechanism for explaining the number of spatial 
dimensions we observe today.
The idea was basically based in the unusual thermodynamical
behaviour of a gas of strings near the Hagedorn temperature 
when considered in a compact space \cite{Brandenberger:1989aj}
(see \cite{Bassett:2003ck,Easther:2004sd,Danos:2004jz} 
for recent reexaminations of this phase).
The string spectrum supports states which can wind around
different directions.
The energy of these modes increases with the size of the 
spatial dimension and this implies that the string gas 
acts as an effective confining potential. 
Then, the Universe can grow large if the winding 
modes are able to decay.
The crucial argument is to note that the annihilation process
of one dimensional objects could only be efficient if the 
dimensionality of the spacetime is smaller than four
\cite{Tseytlin:1992xk,Tseytlin:1992ss}.
The mechanism can be easily extended to include Dp-branes
\cite{Alexander:2000xv}. 
In this case a further hierarchy of small dimensions can also be 
obtained.
To have a better understanding of the physical ideas behind 
this proposal authors have studied two potential implementations
of the mechanism. 
The first in the framework of ten-dimensional dilaton-gravity
\cite{
Brandenberger:2001kj,Campos:2003gj,Easson:2001fy,Easther:2002mi,
Watson:2002nx,Boehm:2002bm,Kaya:2003py,Kaya:2003vj,
Brandenberger:2003ge,Campos:2003ip,Biswas:2003cx,
Watson:2003uw,Watson:2004vs,
Watson:2003gf,Battefeld:2004xw,Kim:2004ca,Watson:2004aq,
Kaya:2004yj,Patil:2004zp,Arapoglu:2004yf,Berndsen:2004tj,
Brandenberger:2005bd}
and the second in eleven-dimensional supergravity
\cite{Easther:2002qk,Alexander:2002gj,Easther:2003dd,Campos:2004yn}.

The purpose of this work is twofold.
On the one hand, we present new cosmological solutions
with fluxes characterised by a large number of unwrapped 
dimensions that can account for the right number of spatial 
dimensions at late times within the context of eleven-dimensional 
supergravity. 
As noticed in \cite{Campos:2004yn} the presence of fluxes
allows the existence of relevant cosmological solutions with 
different numbers of unwrapped dimensions and consequently 
they are capable of softening the fine-tuning problem  
posed by a decompactication mechanism driven by M2-branes
\cite{Easther:2002qk}.
On the other hand, we are interested in understanding to what 
extend the two above mentioned implementations of the 
Brandenberger-Vafa mechanism are connected. 
Are they physically equivalent? 
Can this relation provide a broader perspective of the limits 
of the mechanism and the details of the underlying cosmological 
dynamics?
These questions were first briefly discussed in 
\cite{Easther:2003dd}.
Here we give a more elaborate analysis by incorporating the
late dynamics of gauge fields.

In Sec.~\ref{sec:11D} we derive two new cosmological solutions 
for a M2-brane gas for different numbers of unwrapped dimensions.
Both solutions are driven by a solitonic (magnetic) flux
and complement the solutions with elementary (electric) fluxes
already found in \cite{Campos:2004yn}. 
The first solution we present has eight unwrapped dimensions
which is the largest possibility for a nontrivial anisotropic 
wrapping of the brane gas. 
Configurations with a large number of unwrapped dimensions are 
relevant because they could have been originated in the Hagedorn 
phase from an initial state of the Universe with a small volume.
The second solution has no unwrapped dimensions and is the
M-theory counterpart of the ten-dimensional dilaton-gravity 
solution with a NS flux studied in \cite{Campos:2003ip}.
The cosmological dynamics of this lower dimensional solution is 
analytically discussed in Sec.~\ref{sec:10D} and the relationship 
between both solutions is obtained in Sec.~\ref{sec:up}.
Finally, in Sec.~\ref{sec:con} we outline the conclusions
of our results.

\section{Brane gas dynamics in eleven dimensions}
\label{sec:11D}
\subsection{Field equations}
Let us start considering the gauge-gravity sector of 
eleven-dimensional supergravity, 
\begin{equation}
S  = \frac{1}{2\kappa^2_{11}}
       \int d^{11}x\ \sqrt{-g}
          \left( R - \frac{1}{48} F^2_{[4]}
          \right)\, ,
\end{equation}
where $R$ is the scalar curvature, $F_{[4]}=dA_{[3]}$ is the field 
strength of the gauge field $A_{[3]}$, and the eleven-dimensional 
gravitational coupling constant is given by $2\kappa^2_{11}=(2\pi)^8$ 
in Planck units.
In this work we only consider solutions for the gauge field
with vanishing Chern-Simons contribution.
The classical equations of motion derived from this action
are,
\begin{eqnarray}
G^{\mathrm{IJ}}
   & = &  \kappa^2_{11}
          \left( T^{\mathrm{IJ}}_{_G} + T^{\mathrm{IJ}}_{_M}
          \right) \, ,                                      \label{eq:Einstein}
\\
\nabla_{\mathrm{I}} F^{\mathrm{IJKL}}
   & = & 0\, ,                                               \label{eq:Maxwell}
\end{eqnarray}
where $T^{\mathrm{IJ}}_{_G}$ is the energy-momentum tensor associated with
the gauge field,
\begin{equation}
T^{\mathrm{IJ}}_{_G}
   = \frac{1}{12\kappa^2_{11}} 
     \left( {F^{\mathrm{I}}}_{\mathrm{KLM}}
            F^{\mathrm{JKLM}}
           -\frac{1}{8}g^{\mathrm{IJ}}
            F_{\mathrm{KLMN}}
            F^{\mathrm{KLMN}}
     \right)\, ,
\end{equation}
and $T^{\mathrm{IJ}}_{_M}$ stands for the energy-momentum tensor
of any other matter component.
Here capital Roman indices take values from 0 to 10.
Recall that the field strength must also obey the Bianchi identity 
$\nabla_{[\mathrm{I}} F_{\mathrm{JKLM}]}=0$.

We assume that the spacetime is described by the homogenous
but spatially anisotropic metric,
\begin{equation}
ds^2
   = -dt^2 + \sum^{10}_{i=1} e^{2\lambda_i(t)} dx^2_i\, ,
\end{equation}
with the spatial sections having the topology of a torus,
$0 \leq x_i \leq 1$.
The total volume of this background is given by, 
\begin{equation}
V 
   = \prod^{10}_{i=1}e^{\lambda_i}\, .
\end{equation}
For the gauge field we consider a solution of solitonic 
(magnetic) type. 
In this solution the 4-form field strength is assumed 
to be nonzero only in a four-dimensional spatial 
submanifold,
\begin{equation}
F^{\mathrm{IJKL}}
   = \frac{\epsilon^{\mathrm{IJKL}}}{\sqrt{g_4}} F(t)\, ,
\end{equation}
where $g_4$ is the determinant of the induced metric on 
the submanifold and $\epsilon^{\mathrm{IJKL}}$ is the 
corresponding Levi-Civit\`a density.
When indices run, for instance, from 7 to 10, the function 
$F(t)$ is simply given by,
\begin{equation}
F(t)
   = f\, 
     e^{-\lambda_7(t)}\cdots e^{-\lambda_{10}(t)}\, ,      \label{eq:solitonic}
\end{equation}
with $f$ a constant of integration.
The energy-momentum tensor for the gauge field is diagonal 
and corresponds to a fluid with energy density,
\begin{equation}
\rho_{_G} 
   = \frac{1}{\kappa^2_{11}}
           \left( \frac{F(t)}{2}
           \right)^2\, ,                               \label{eq:rho_solitonic}
\end{equation}
and anisotropic pressures,
\begin{equation}
p^i_{_G} 
   = \varepsilon^i \rho_{_G}\, ,                         \label{eq:p_solitonic}
\end{equation}
where the ten-dimensional object $\varepsilon^i$ indicates
a plus or minus sign. 
For the solution explicitly indicated in (\ref{eq:solitonic}),
it is $-1$ for $i=1, \cdots 6$ and $+1$ for $i=7, \cdots, 10$.

We further  assume that an important component of supersymmetric
relativistic matter is present in the early Universe.
This source can be represented by a gas of massless particles, 
with energy density $\rho_{_S}$ and pressure $p_{_S}$. 
For simplicity we take the gas to be a homogeneous and isotropic
perfect fluid with a radiation equation of state 
$p_{_S} = \rho_{_S}/10$.
Note that because the energy-momentum tensor is individually
covariantly conserved, the energy density of supergravity 
particles scales with the total size of the Universe as,
\begin{equation}
\rho_{_S}
   = \rho^o_{_S} \left( \frac{V_o}{V}
                 \right)^{11/10}\, ,
\end{equation}
where $\rho^o_{_S}$ and $V_o$ are, respectively, the energy 
density and the spatial volume at some given time, $t_o$.

The last source of matter in this model is a gas of 
M2-branes wrapping the various cycles of the torus.
The possible brane wrappings can be characterised by a symmetric 
matrix $N_{ij}$, where elements with $i<j$ ($i>j$) represent the 
number of branes (anti-branes) wrapped around the cycle $(ij)$.
This effective description separates the spatial dimensions into 
three groups.
A direction is said to be {\sl unwrapped} if it has no branes at all.
A {\sl Fully wrapped} direction is that which shares branes
with all the dimensions but the unwrapped ones. 
The rest of directions are called {\sl partially wrapped}.
Accordingly, we will refer to a brane gas with $m_1$ unwrapped,
$m_2$ partially wrapped, and $m_3$ fully wrapped dimensions as
a brane gas of wrapping type $m_1$-$m_2$-$m_3$.
To study the late time dynamics we start at the point in which
the brane gas has already freezed out and the elements of 
the wrapping matrix are constants.
In principle, the actual values will depend on the thermodynamics
of the gas in the Hagedorn phase but for practical purposes we
can take them as random integer numbers.
Ignoring excitations on the brane and assuming that the branes
are nonrelativistic, an energy-momentum tensor properly smeared 
over transverse directions can be defined for the gas.
The corresponding energy density and pressures being, respectively,
\begin{eqnarray}
\rho_{_B} 
   & = &  \frac{T_2}{V} 
          \sum_{k \neq l} e^{\lambda_k} e^{\lambda_l} N_{kl}\, , 
\\
p^i_{_B} 
   & = & - \frac{T_2}{V} 
           \sum_{k \neq i} e^{\lambda_i} e^{\lambda_k}
           \left( N_{ki} + N_{ik}
           \right)\, ,
\end{eqnarray}
where $T_2=1/(2\pi)^2$ is the brane tension.

Inserting all the matter sources in the Einstein 
equations~(\ref{eq:Einstein}) one gets the Friedmann 
constraint, 
\begin{equation}
\sum^{10}_{k\neq l} \dot\lambda_k\dot\lambda_l
   =
    2\kappa^2_{11}
     \left( \rho_{_S} + \rho_{_B} + \rho_{_G}
     \right)\, ,                                          \label{eq:Einstein_t}
\end{equation}
and dynamical equations for the sizes of all the spatial
dimensions,
\begin{equation}
\ddot\lambda_i + 10 H \dot\lambda_i
   = \kappa^2_{11}
     \left[ \frac{1}{10}\rho_{_S}
           +\frac{1}{3}\rho_{_B}
           +\left( \varepsilon^i
                  +\frac{1}{3} 
            \right) \rho_{_G}
           +p^i_{_B}
     \right]\, ,                                          \label{eq:Einstein_s}
\end{equation}
with $i=1, \cdots, 10$.
Here, $H$ is the mean Hubble function which represents the 
rate change of the total spatial volume of the Universe and is
given in terms of the metric components by,
\begin{equation}
H
   = \frac{1}{10}\sum^{10}_{i=1}\dot\lambda_i\, .
\end{equation}
It is important to notice that in (\ref{eq:Einstein_s}) there is 
a change of sign in the prefactor of the energy density for the 
gauge field with respect to that on the equations for an elementary 
type solution \cite{Campos:2004yn}.

Looking into Eq.~(\ref{eq:Einstein_s}), one can readily 
observe that there are two possible sources of anisotropy 
in the cosmological dynamics.
The one coming from the brane gas $p^i_{_B}$ and the one 
from the gauge field $\varepsilon^i \rho_{_G}$. 
As long as those matter components have a significant
contribution to the total expansion at late times the
differential pressures will allow an asymmetric evolution 
on the sizes of different dimensions. 
For instance, consider the situation in which the gauge
field is the only dominant matter component and is described
by the solution given in (\ref{eq:solitonic}).
At late times, a scaling solution to the Einstein equations 
can be found with, 
\begin{equation}
e^{\lambda_i}
   \sim \left\{
               \begin{array}{ll}
                 t^{-1/8} & \textrm{\,\,\, for } i=1, \cdots, 6 \\
                 t^{1/4} & \textrm{\,\,\, for } i=7, \cdots, 10
               \end{array}
         \right.                                          \label{eq:noM2branes}
\end{equation}
This means that the effect of a solitonic flux solution is 
to drive the Universe with 4 expanding and 6 contracting
dimensions. 
Then, the gauge field naturally splits the spacetime into 
R$\times$T$^6\times$T$^4$ and the proper number of large 
spatial dimensions can not be explained.
In the next sections we will see in some particular cases
how this behaviour affects a potential mechanism for 
explaining the actual number of spatial dimensions based on 
the dynamics of a gas of M2-branes. 

\subsection{Cosmological solution with 8 unwrapped dimensions}
It was shown in \cite{Campos:2004yn} that the presence 
of an elementary flux can support the grow of three large 
directions at late times for a brane gas configuration with 
6 unwrapped dimensions.
Here, we present a new asymptotic solution for a configuration 
with 8 unwrapped dimensions which is the maximum nontrivial 
number of unwrapped dimensions after freeze out.
Note that due to the spatial dimensionality of the M2-branes
there cannot exist 9 unwrapped dimensions and that an evolution
with 10 unwrapped dimensions (no brane left after freeze out)
is always isotropic.
This new solution is characterised by a brane gas with wrapping 
8-0-2 (that is, a gas configuration with eight unwrapped and two 
fully wrapped dimensions) and is driven by a solitonic flux of the 
form (\ref{eq:solitonic}). 

Without a gauge field the asymptotic cosmological dynamics of a 
brane gas with general wrapping $m_1$-$m_2$-$m_3$ can be cast by 
a scaling solution of the form \cite{Easther:2002qk},
\begin{equation}
e^{\lambda_i}
   \sim \left\{
               \begin{array}{ll}
                 t^a & \textrm{\,\,\, for } i=1, \cdots, m_1 \\
                 t^b & \textrm{\,\,\, for } i=m_1+1, \cdots, m_1+m_2\\
                 t^c & \textrm{\,\,\, for } i=m_1+m_2 \cdots, m_1+m_2+m_3 
               \end{array}
         \right.                                          \label{eq:no-fluxes}
\end{equation}
When $m_2\leq m_3$ and $5\leq m_1 < 9$, both the gas of supergravity
particles and the gas of branes are not negligible at late times.
In this case the generic solution only depends on the number of
unwrapped dimensions $m_1$ and is explicitly given by,
\begin{eqnarray}
a
   & = & \frac{30-m_1}{11m_1}\, ,
\nonumber\\
b
   & = &
c
     =   -\frac{1}{11}\, .
\nonumber
\end{eqnarray}
Then, for a brane gas with a wrapping configuration of type 
8-0-2 one simply has,
\begin{equation}
a  = \frac{1}{4}, \,\,\,\,
c  = -\frac{1}{11}\, ,                                         \label{eq:8-0-2}
\end{equation}
and, subsequently, the eight unwrapped dimensions are expanding 
and the two fully wrapped dimensions are contracting.

\begin{figure}[t]
\includegraphics*[totalheight=\columnwidth,width=\columnwidth]{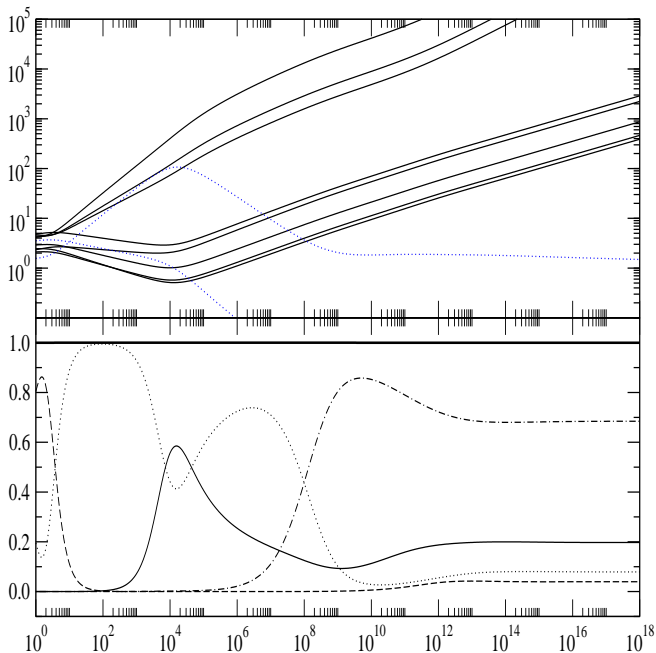}
\caption{Cosmological evolution for a brane gas of wrapping 
type 8-0-2 with a solitonic flux. {\sl Top:} Sizes of the spatial 
dimensions as functions of cosmic time.  Solid curves are for unwrapped 
dimensions, dashed curves for partially wrapped dimensions, and dotted 
curves for fully wrapped dimensions. {\sl Bottom:} Evolution of the 
different fractional contributions to the total expansion of the 
Universe. The solid line represents the contribution from the energy 
density of the M2-brane gas, the dashed line from the energy density 
of the gauge field, the dotted-dashed line from the energy density of 
supergravity particles, and the dotted line is the fractional shear 
component. The thick solid line is the sum of all the contributions and 
serves to check the accuracy of the numerical computation.   
\label{fig:8-0-2-flux}}
\end{figure}

Let us now analyse the dynamical effects of the gauge field.
First note that for a brane configuration of type 8-0-2 the
only nonvanishing component of the wrapping matrix is 
$N_{9\, 10}$.
This readily implies that the individual pressures of the 
brane gas are,
\begin{eqnarray}
p^i_{_B}  
   & = & 0\textrm{\,\,\, for } i=1, \cdots, 8\, , 
\nonumber\\
p^9_{_B}  
   & = & p^{10}_{_B}  = -\rho_{_B}\, .
\nonumber
\end{eqnarray} 
For concreteness consider the solitonic flux picks coordinates 
$1, 2, 3$ and $10$. 
(Note that other possible combinations also lead to similar 
results with three large dimensions at late times.
The important point is that the flux takes three unwrapped 
dimensions.)
The presence of the gauge field introduces a new physical
length scale, $l_{_G}$ into the problem.
Following similar arguments as those in \cite{Campos:2004yn}
for the elementary case, one can estimate this scale to be
of the order $l_{_G}\sim l^4_o f^{-1}$ where $l_o$ represents
the typical length scale for the initial size of the Universe.  
The gauge field will has a relevant contribution to the dynamics 
as long as the field strength has an integration constant of 
the order of $f\sim l^{-3}_o$.
A full numerical solution of the cosmological dynamics
fulfilling such condition is shown in Fig.~\ref{fig:8-0-2-flux}. 
The top plot represents the evolution of the size of each
spatial dimension and the bottom plot the evolution of
all fractional contributions to the total expansion  
of the Universe, including the fractional shear component 
which gives a measure of the amount of anisotropy of the 
spacetime (for the actual definition in our context see
\cite{Campos:2004yn}).

As one can observe, in this solution with eight unwrapped 
dimension, three spatial dimensions can get large dynamically.
The energy density of the gauge field starts dominating the
evolution and, consequently, four dimensions are forced to
expand and six to contract initially (recall 
Eq.~(\ref{eq:noM2branes})).
As the size of the four expanding dimensions grows, the
energy density of the gauge field (\ref{eq:rho_solitonic}) 
decreases.
This behaviour continues until the energy density of the
brane gas takes over the rest of energy components causing
the Universe to evolve with eight expanding and two contracting
dimensions. 
After that turning point there appears a transient phase 
in which the energy density of the brane gas decreases and
the energy density of the supergravity particles increases.
This period is characterised by a fast drop of the total amount 
of anisotropy. 
The final asymmetry in the size of dimensions occurs at the
time when all the energy densities stabilise.

The asymptotic behaviour of this solution can be studied 
analytically using a scaling ansatz of the form,
\begin{equation}
e^{\lambda_i}
   \sim \left\{
               \begin{array}{ll}
                 t^{a_+} & \textrm{\,\,\, for } i=1, \cdots, 3 \\
                 t^{a_-} & \textrm{\,\,\, for } i=4, \cdots, 8 \\
                 t^{c_-} & \textrm{\,\,\, for } i=9            \\
                 t^{c_+} & \textrm{\,\,\, for } i=10 
               \end{array}
         \right.
\end{equation}
The $\pm$ signs in the exponents have been introduced to 
keep track of the dimensions where the field strength of the
gauge field is nonzero.
As we have seen, the three matter components have an important
fractional contribution to the expansion at late times.
Imposing that all the energy densities scale with cosmic time
as $t^{-2}$ and demanding selfconsistency of the Einstein equations
a set of four algebraic relations can be found,
\begin{eqnarray}
  a_- - a_+ - c_- + c_+ & = & 0\, ,\nonumber\\
5a_- + 3a_+ + c_- + c_+ & = & \frac{20}{11}\, ,\nonumber\\
             3a_+ + c_+ & = & 1\, ,\nonumber\\
            5a_- + 3a_+ & = & 2\, .\nonumber
\end{eqnarray}
Solving for the exponents one obtains,
\begin{eqnarray}
&& a_+ = \frac{71}{209}, \,\,\,\, a_- = \frac{41}{209}\, ,\nonumber\\
&& c_- = -\frac{34}{209}, \,\,\,\, c_+ = -\frac{4}{209}\, ,\nonumber
\end{eqnarray}
which can be compared with the numerical result,
\begin{eqnarray}
&& a_+ = 0.33972, \,\,\,\, a_- = 0.19614\, ,\nonumber\\
&& c_- = -0.1628, \,\,\,\, c_+ = -0.019222\, .\nonumber
\end{eqnarray}

To sum up, {\sl we have found a cosmological solution driven
by a brane gas of M2-branes and a solitonic flux leading to
the appropriate asymmetry among spatial dimensions with the
largest number of nontrivial unwrapped dimensions allowed}.
Brane configurations with a large number of unwrapped 
dimensions can be originated after freeze out from a smaller
initial size of the Universe than configurations with lower
unwrapping numbers \cite{Easther:2003dd}. 
In addition to the above solution we have check that other 
relevant solutions can also be obtain with seven and zero
unwrapped dimensions.
In the next sections we study a particular solution with 
zero unwrapped dimensions which reflects a connection
between the cosmology of a brane gas in eleven dimensions
and that of a string gas in ten dimensions.

\subsection{Cosmological solution with wrapping 0-9-1}
Now we are interested to study the asymptotic cosmological 
dynamics of a brane gas characterised by a wrapping of type
0-9-1 with a solitonic flux. 
As we will see later, this solution is the higher dimensional 
counterpart of the cosmological solution found in 
\cite{Campos:2003ip} for a particular ten-dimensional
dilaton-gravity implementation of the Brandenberger-Vafa 
mechanism. 

Without a gauge field, a gas of M2-branes described by a wrapping 
configuration that obeys the conditions $m_1=0$, $0<m_3<m_2$, 
and $2m_2m_3<25$, has the generic asymptotic power law solution 
(\ref{eq:no-fluxes}) with,
\begin{eqnarray}
b  & = &  \frac{2}{11}\frac{m_3+10}{m_2-m_3}\,\nonumber\\
c  & = & -\frac{2}{11}\frac{m_2+10}{m_2-m_3}\,\nonumber\\
\end{eqnarray}
Substituting for our particular case with $m_2=9$ and $m_3=1$ 
one obtains,
\begin{equation}
b  = \frac{1}{4}, \,\,\,\,
c  = -\frac{19}{44}\, .                                        \label{eq:0-9-1}
\end{equation}
As expected, this configuration has nine dimensions which 
are expanding and one which is contracting.
In this case both the energy density of the brane gas and the 
energy density of supergravity particles have a relevant 
contribution to the total expansion of the Universe at late times. 

\begin{figure}[t]
\includegraphics*[totalheight=\columnwidth,width=\columnwidth]{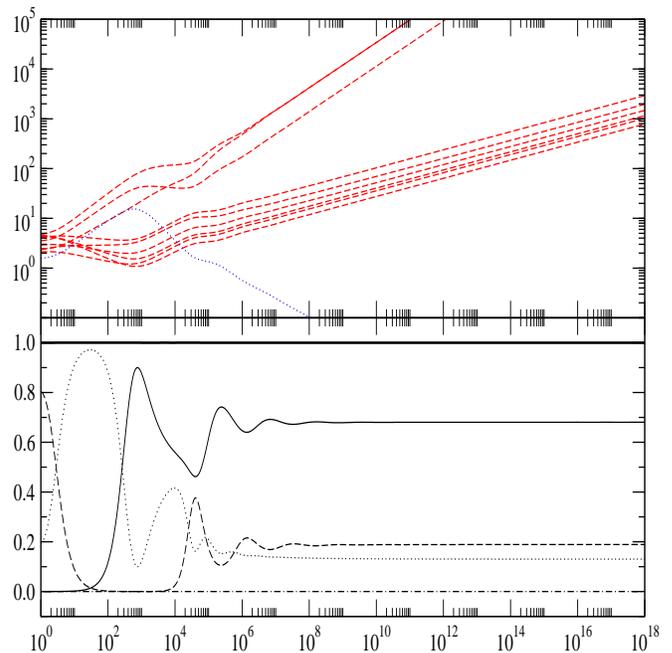}
\caption{Cosmological evolution for a brane gas of wrapping 
type 0-9-1 with a solitonic flux. The identification of plots and lines 
is the same as in Fig.~\ref{fig:8-0-2-flux}.
\label{fig:0-9-1-flux}}
\end{figure}

Let us now analyse the cosmological evolution of this
brane configuration including the gauge degrees of freedom.
The field strength is assumed to lay in the submanifold 
parametrised by the last four coordinates.
Again the late attractor solution can be cast with an ansatz
of the form,
\begin{equation}
e^{\lambda_i}
   \sim \left\{
               \begin{array}{ll}
                 t^{b_-} & \textrm{\,\,\, for } i=1, \cdots, 6 \\
                 t^{b_+} & \textrm{\,\,\, for } i=7, 8, 9 \\
                 t^{c_+} & \textrm{\,\,\, for } i=10
               \end{array}
         \right.                                          \label{eq:0-9-1sol.a}
\end{equation}
The numerical solution of the full equations has been
plotted in Fig.~\ref{fig:0-9-1-flux}.
As one can observe, at late times only the contribution of
the brane gas and the gauge field are relevant.
This occurs as long as the energy density of both components
scale with cosmic time as $t^{-2}$.
Using the definition of these energy densities one finds two
algebraic conditions,
\begin{eqnarray}
3{b_-}+{b_+} &=& 1\, ,
\nonumber\\
3{b_+}+{c_+} &=& 1\, .
\nonumber
\end{eqnarray}
On the other hand, the spatial components of Einstein equations
(\ref{eq:Einstein_s}) impose another additional relation,
\begin{equation}
2{b_-}+{c_+} = 0\, .
\end{equation}
Solving for ${b_+}, {b_-}$ and ${c_+}$ the linear system of three 
equations, one obtains,
\begin{equation}
{b_+}  = \frac{5}{11}, \,\,\,\,
{b_-}  = \frac{2}{11}, \,\,\,\,
{c_+}  = -\frac{4}{11}\, .                                \label{eq:0-9-1sol.b}
\end{equation}
In our numerical solution these parameters are obtained with
an error of one part in $10^4$.
Note that for this configuration the appropriate asymmetry of 
dimensions to explain the dimensionality of spacetime is also
achieved.
One direction is contracting and nine expanding with three growing
more rapidly than the other six.

\section{String gas dynamics in ten dimensional dilaton-gravity}
\label{sec:10D}
As we will see shortly the previous eleven-dimensional 
supergravity solution with wrapping type 0-9-1 is deeply 
connected with another solution in the ten-dimensional 
dilaton-gravity scenario for the implementation of the 
Brandenberger-Vafa mechanism.
In particular, it is related to the string/gauge solution 
presented in \cite{Campos:2003ip}.
In that solution the cosmological dynamics is driven by 
a freezed gas of strings filling all the dimensions and a 
solitonic two-form gauge field.
Without the dynamical effects of the flux the evolution of 
the string gas is completely isotropic but when it is taking 
into account three spatial dimensions can get large and six
stabilise naturally.
For simplicity, this was shown numerically for a background 
with three spatial dimensions.
Here we complement the analysis of this solution with an 
analytical discussion in the appropriate number of 
dimensions.

\subsection{Set-up}
We start with the ten-dimensional bosonic action with fields
only in the NS-NS sector,
\begin{equation}
S  = \frac{1}{2\kappa^2_{10}}
       \int d^{10}x\ \sqrt{-g}
          e^{-2\phi}
          \left( R 
                +4(\nabla_\mu\phi)^2
                - \frac{1}{12} H^2_{\alpha\beta\gamma}
          \right)\, ,
\end{equation} 
where $\phi$ represents the dilaton field, $R$ is the 
scalar curvature of the ten-dimensional metric, and
$H_{\alpha\beta\gamma}$ the field strength of the two-form 
gauge field.
(Greek indices run from 0 to 9.)
The field strength obeys two equations,
\begin{eqnarray}
\nabla_\alpha ( e^{-2\phi} H^{\alpha\beta\gamma})
   & = & 0\, ,
\\
\nabla_{[\beta} H_{\mu\nu\alpha]}
   & = & 0\, .
\end{eqnarray}
The first comes as a result of the variation of the
action with respect to the gauge degrees of freedom
and the second is the Bianchi identity.
Following \cite{Campos:2003ip}, we consider a solitonic 
solution in which the components of the field strength 
are restricted to a three-dimensional spatial submanifold.
By adopting a homogenous metric of the form,
\begin{equation}
ds^2
   = -N^2(t)dt^2 
     +e^{2\nu(t)} \sum^{6}_{i=1} dx^2_i
     +e^{2\lambda(t)} \sum^{9}_{i=7} dx^2_i\, ,            \label{eq:metric-10}
\end{equation}
the appropriate ansatz for the flux is,
\begin{equation}
H^{\alpha\beta\gamma}
   = \frac{\epsilon^{\alpha\beta\gamma}}{\sqrt{g_3}} H(t)\, ,
\end{equation}
where, analogously to the eleven-dimensional ansatz, $g_3$ 
is the determinant of the induced three-dimensional metric 
and $\epsilon^{\alpha\beta\gamma}$ is the Levi-Civit\`a 
antisymmetric density.
In order to make a direct connection to the higher dimensional 
brane solution of wrapping 0-9-1, we assume the indeces 
$\alpha, \beta$, and $\gamma$ take values 7, 8, and 9. 
The above ansatz obeys the dynamical equation identically
and the function $H(t)$ is actually fixed by the Bianchi 
identity, 
\begin{equation}
H(t) 
   = H_o e^{-3\lambda}\, ,
\end{equation}
where $H_o$ is a constant of integration.
This simple solution yields a gauge term on the action that 
only depends on $\lambda$,
\begin{equation}
H^2_{\alpha\beta\gamma}
   = 6 H^2_o e^{-6\lambda}
   \equiv 12U(\lambda)\, .
\end{equation}

For this solution of the gauge field, the bosonic action 
can be rewritten as,
\begin{equation}
S  = \frac{1}{2\kappa^2_{10}}
       \int dt
          e^{-\varphi}
          \left[ \frac{1}{N}
                 \left( 3\dot\lambda^2
                       +6\dot\nu^2
                       - \dot\varphi^2
                 \right)
                - N U(\lambda)
          \right]\, ,
\end{equation}
where we have introduced the shifted dilaton variable
$\varphi\equiv 2\phi - 3\lambda - 6\nu$ in order to 
absorb the space volume factor.
Taking the gauge $N=1$ and units in which $2\kappa^2_{10}=1$, 
the equations of motion are,
\begin{eqnarray}
\dot\varphi^2 - 3\dot\lambda^2 - 6\dot\nu^2
   & = & e^{\varphi} E + U(\lambda)\, ,
\\
\ddot\varphi - 3\dot\lambda^2 - 6\dot\nu^2
   & = & \frac{1}{2} e^{\varphi} E\, ,
\\
\ddot\lambda - \dot\varphi\dot\lambda
   & = & \frac{1}{2} e^{\varphi} P_\lambda + U(\lambda)\, ,
\\
\ddot\nu - \dot\varphi\dot\nu
   & = & \frac{1}{2} e^{\varphi} P_\nu\, .
\end{eqnarray}
In the above equations $E$ denotes the total energy
of the string gas and, $P_\lambda$ and $P_\nu$ the
total pressures in the three- and six-dimensional
subvolumes, respectively.
We are interested in a situation in which 
the spacetime background is still effectively higher 
dimensional but the strings present do not longer 
interact.
The interaction ceases when the annihilation
rate is suppressed with respect to the expansion 
rate. 
We assume that freeze-out occurs sufficiently
fast and then the remaining network of strings 
wraps all the nine spatial dimensions.  
Such initial configurations are supported by recent
studies of the string thermodynamics close to the
Hagedorn phase \cite{Easther:2004sd,Danos:2004jz}.
In this case, the total energy of the string 
gas, $E$, can be separated into the contribution 
from the $T^3$ subvolume, $E_\lambda$, and that of
the $T^6$ subvolume, $E_\nu$,
\begin{equation}
E
   = E_\lambda + E_\nu\, .
\end{equation}
For strings in a compact space these individual energies 
can be expressed as,
\begin{eqnarray}
E_\lambda
   & = & \mu N_{(3)} e^\lambda\, ,
\\
E_\nu
   & = & \mu N_{(6)} e^\nu\, ,
\end{eqnarray}
where $\mu$ is a mass parameter, and $N_{(3)}$ and $N_{(6)}$ 
account for the number of winding modes in the corresponding 
subvolume.
On the other hand, the total pressure of the string gas
in each subvolume are related with the above energies by,
\begin{eqnarray}
P_\lambda
   & = & -\frac{1}{3} E_\lambda\, ,
\\
P_\nu
   & = & -\frac{1}{6} E_\nu\, .
\end{eqnarray}
Recall that, in general, for a nonrelativistic gas of Dp-branes 
in a $d$-dimensional space the total pressure is given in terms
of the energy by $P=-(p/d)E$.

\subsection{Late-time dynamics}
Without gauge fields the above string gas set-up will 
certainly lead to a contracting Universe with a unique scale 
factor scaling as $t^{-1/5}$ at late times and consequently no 
asymmetry among different dimensions.
Consider the opposite situation in which the energy density of 
the string gas is negligible.
In this case the scale factors of the two subvolumes behave as,
\begin{eqnarray}
e^\lambda
   & \sim & t^{\sqrt{3}/3}\, ,
\\
e^\nu
   & \sim & {\mathrm constant}\, .
\end{eqnarray}
Although a large Universe with three dimensions expanding 
and six dimensions stabilised seems to be explained, one can 
easily see that the theory becomes strongly coupled and, then, 
the low energy description used to describe the dynamics is
inappropriate \cite{Campos:2003ip}.

The interesting situation occurs when both the string gas and
the gauge field have a significant contribution to the total
expansion of the Universe at late times.
Let us see that the evolution presents an attractor power-law 
behaviour asymptotically.
One starts by taking all fields scaling with cosmic time as, 
\begin{equation}
e^\lambda
   \sim t^\alpha, \,\,\,\,
e^\nu
   \sim t^\beta, \,\,\,\,
e^\varphi
   \sim t^\gamma\, .                                     \label{eq:dilaton_sol}
\end{equation}
To determine $\alpha$ and $\gamma$ one simply needs to look
into the equation of motion for the variable $\lambda$.
The right hand side of this equation can be interpreted as 
coming from a classical effective potential,
\begin{equation}
V_{eff}(\lambda)
   = \frac{1}{6}
     \left[ \mu N_{(3)} e^\varphi e^\lambda
           + U(\lambda)
     \right]\, .
\end{equation}
The first term comes from the string gas and produces
a confinement force whereas the second term is the contribution
from the gauge field and induces an expanding force.
In an expanding Universe one would naively expect that the gauge
contribution should always be less important than the string gas 
contribution.
However, this does not need to be the case because the dilaton 
dynamics is modulating the first term in the effective potential.
This means that both contributions can only be important at late
times simultaneously if $\varphi$ is a decreasing function of time.
This property is guaranteed by the equations of motion as long as 
$\dot\varphi$ is initially negative.
Imposing that both terms in the effective potential scale 
asymptotically with cosmic time as $t^{-2}$ immediately yields,
\begin{equation}
\alpha 
   = \frac{1}{3}, \,\,\,\,
\gamma
   = -\frac{7}{3}\, .                                         \label{eq:10Dsol}
\end{equation}
On the other hand, the consistency of the dynamical equation for 
$\nu$ additionally gives,
 \begin{equation}
\beta 
   = 0\, ,
\end{equation}
which means that the directions that do not get large are stabilised
at a constant radius.
This is easily understood by examining the effective potential
for the field $\nu$,
\begin{equation}
V_{eff}(\nu)
   = \frac{1}{12}
     \mu N_{(6)} e^\varphi e^\nu\, ,
\end{equation}
which represents a confining exponential force.
The evolution of $\nu$ is stopped simply because the shape
of the potential is completely flatten out by the dilaton field 
dynamics.
Finally, note that with these parameters the original dilaton field
behaves as $e^\phi\sim t^{-2/3}$ and then the weak coupling condition 
is dynamically preserved by this analytical solution.

Then, as a conclusion, the effect of the gauge field at late 
times in this string gas scenario is {\sl to make the size of 
three dimensions large and at the same time to stabilise the 
small extra dimensions} giving a simple realization of the 
Brandenberger-Vafa mechanism. 

\section{Uplifting string gas solutions to eleven dimensions}
\label{sec:up}
Uplifting cosmological solutions from string gases in ten 
dimensions to eleven dimensions was briefly discussed in 
\cite{Easther:2002qk}.
Here, we extend their discussion by considering solutions
with gauge fields.

Consider the solution studied in the previous section.
Asymptotically, it can be described by the ten-dimensional
metric, 
\begin{equation}
ds^2_{10}
   = -dT^2
     +\sum^{6}_{i=1} dy^2_i
     +T^{2\alpha} \sum^{9}_{i=7} dy^2_i\, ,               \label{eq:10d_metric}
\end{equation}
and a late evolution for the shifted dilaton field,
\begin{equation}
e^{\varphi}
   \sim T^\gamma\, ,
\end{equation}
where $\alpha$ and $\gamma$ are given by (\ref{eq:10Dsol}).
In general, the standard procedure to uplift the ten-dimensional 
solution to one extra dimension is to define the eleven-dimensional
metric,
\begin{equation}
ds^2_{11}
   =  e^{-2\phi/3}\, ds^2_{10}
     +e^{4\phi/3}\, dy^2_{10}\, ,
\end{equation}
where $y_{10}$ represents the tenth spatial coordinate in the 
higher dimensional space and $\phi$
is the original dilaton.
Then, for the asymptotic cosmological solution (\ref{eq:10d_metric}),
the eleven-dimensional metric should be,
\begin{equation}
ds^2_{11}
   =  T^{-2\bar\gamma}\, ds^2_{10}
     +T^{4\bar\gamma}\, dy^2_{10}\, ,
\end{equation}
with,
\begin{equation}
\bar\gamma
   = \frac{1}{6}\left( \gamma + 3\alpha
                \right)
   = - \frac{2}{9}\, .
\end{equation}

To understand the physical meaning of this solution
one has to write the metric with respect to proper time.
Defining a new time coordinate $t$ as,
\begin{equation}
t
   = \frac{T^{1-\bar\gamma}}{1-\bar\gamma} \, ,
\end{equation}
and introducing new spatial coordinates $x_i$, in order to 
absorb irrelevant constant factors, the higher dimensional
metric can be rewritten as,
\begin{equation}
ds^2_{11}
   = -dt^2
     +t^{-\frac{2\bar\gamma}{1-\bar\gamma}}\,
      \sum^{6}_{i=1} dx^2_i
     +t^{\frac{2(\alpha-\bar\gamma)}{(1-\bar\gamma)}}\, 
      \sum^{9}_{i=7} dx^2_i
     +t^{\frac{4\bar\gamma}{1-\bar\gamma}}\, 
      dx^2_{10}\, .
\end{equation}

It is straightforward to see, by substituting the value of the 
parameters, that this metric corresponds to the asymptotic 
solution (\ref{eq:0-9-1sol.a}) and (\ref{eq:0-9-1sol.b}) 
described in Sec.~\ref{sec:11D}.
This relationship permits a physical eleven-dimensional 
interpretation of the ten-dimensional solution.
{}From the higher dimensional perspective, instead of a 
gas of strings covering the whole space with a solitonic 
three-form field strength, one has a M2-brane gas of 
anisotropic wrapping 0-9-1 with a four-form field strength. 
Note that the M2-brane configuration is such that all branes 
have one spatial dimension passing through the tenth direction
which is the one that becomes compact. 
This is the reason why the gas of branes behaves effectively as 
a gas of one-dimensional objects filling the whole nine-dimensional
space in the lower-dimensional gravi-dilaton implementation. 

The important difference between both realizations is that from the 
higher dimensional point of view the compactification of the tenth 
dimension is obtained dynamically and not assumed a priori as being
small.
On the other hand, it is also worth noticing that the small 
dimensions that are stabilised in the dilaton-gravity context are 
destabilised in the M theory framework.
Due to the compactification procedure this is a quite general
property that makes a hard task to obtain stabilisation of the 
extra dimensions in the higher dimensional realization. 
Nevertheless, one can still imagine potential configurations 
that can succeed in stabilising this extra dimensions by
uplifting a particular ten-dimensional solution.
For instance, one can think of a set up in which the momentum 
modes are confined to a three dimensional subspace and a freezed 
gas of strings to the complementary six dimensional subspace.
For this configuration one expects an asymptotic evolution
with three expanding and six contracting dimensions.
After uplifting to eleven dimensions the six contracting spatial 
dimensions can be stabilised if the dynamical evolution 
(\ref{eq:dilaton_sol}) is such that the scaling behaviour of the 
shifted dilaton, $\gamma$, and of the large expanding dimensions,
$\alpha$, obey the condition,
\begin{equation}
\gamma
   = -3\alpha\, .
\end{equation}
{}From the eleven-dimensional perspective three dimensions will 
be expanding, six will be stabilised and one, that related to 
the dilaton, will be contracting.
Now one can ask whether a M2-brane wrapping configuration can be 
constructed with six, or seven if one is not interested to 
connect with a lower dimensional solution, asymptotically stable 
dimensions. 
To answer this question positively what seems unavoidable is the 
necessity of having a nonisotropic distribution of supergravity 
particles.

\section{Conclusions}
\label{sec:con}
This work have been devoted to present new asymptotic cosmological
solutions with solitonic fluxes that could potentially explain 
the dimensionality of the spacetime in both the eleven-dimensional
supergravity and the ten-dimensional dilaton-gravity frameworks. 

In the higher dimensional context we have studied solutions
supporting brane gas configurations with different numbers of
unwrapped dimensions.
This gives stronger evidence to the importance of fluxes, solitonic
or elementary, in order to understand the full space of nontrivial 
solutions in brane gas cosmology. 
It is worth to mention that one of the solutions has eight unwrapped 
dimensions which is the largest possible configuration that can 
drive a nonisotropic cosmological evolution.
This type of brane wrappings could have been produced after freeze
out from a small initial volume of the Universe
\cite{Easther:2003dd}.
In addition, we have also seen that solitonic and elementary
fluxes introduce two different physical length scales and this will 
certainly be important to understand the string/brane thermodynamics 
of the Hagedorn phase.    

We have also illustrated an example in which the lower dimensional
dilaton-gravity solution is related with the M theory solution.  
It would be interesting to investigate {\sl whether any string
gas solution in ten dimensions will have a M2-brane counterpart
in eleven dimensions}. 
One example that could be worth analysing is the string gas 
stabilisation mechanism proposed in \cite{Watson:2003gf}
and find out if it has a brane gas counterpart in the eleven
dimensional implementation of the Brandenberger-Vafa
mechanism.

Finally, It would certainly be interesting to study the 
potential connection of the flux solutions we have found in 
the context of brane gas cosmology and the presence of 
S-branes \cite{Gutperle:2002ai,Chen:2002yq}.

\acknowledgments

The author thanks M. Seco for valuable discussions on the numerical
approach and also acknowledges the support of the Alexander von 
Humboldt Stiftung/Foundation and the Universit\"at Heidelberg.

\bibliography{../../../BIBLIOGRAPHY/bgc}

\end{document}